\documentclass{article}

\input{tcilatex}
\begin{document}

\begin{center}
{\Large Second Order Thermal Corrections to Electron Wavefunction}

\bigskip Mahnaz Q. Haseeb$^{a^{\ast }}$ and\ Samina S. Masood$^{b}$

$^{a}$\textit{Department of Physics, COMSATS Institute of Information
Technology, Islamabad, Pakistan,}

$^{b}$\textit{Department of Physics, University of Houston Clear Lake,
Houston TX 77058}

$_{^{\ast }mahnazhaseeb@comsats.edu.pk}$

\textbf{Abstract}
\end{center}

{\small Second order perturbative corrections to electron wavefunction are
calculated here at generalized temperature, for the first time. This
calculation is important to prove the renormalizeability of QED through
order by order cancellation of singularities at higher order. This
renormalized wavefunction could be used to calculate the particle processes
in the extremely hot systems such as the very early universe and the stellar
cores. We have to re-write the second order thermal correction to electron
mass in a convenient way to be able to calculate the wavefunction
renormalization constant. A procedure for integrations of hot loop momenta
before the cold loop momenta integration is maintained throughout to be able
to remove hot singularities in an appropriate way. Our results, not only
includes the intermediate temperatures $T\sim m$\ (where $m$\ is the
electron mass), the limits of high temperature $T>>m$\ and low temperature $%
T<<m$\ are also retrievable. A comparison is also done with the existing
results. }

PACS: 11.10.Wx, 12.20.-m, 11.10.Gh, 14.60.Cd

Keywords: renormalization, finite temperature field theory, electron self
energy, two-loop QED corrections

\section{Introduction\qquad \qquad}

Finite temperature effects are important in extremely high temperature
environments, such as in the early Universe a few seconds after the Big
Bang, astrophysical environments etc., where they are significant enough and
can not be ignored in comparison with the vacuum contribution. The high
temperature and density effects in ultra-relativistic plasma need to be
incorporated, for example, in Quantum Electrodynamics (QED) plasma, quark
gluon plasma and the core of dense stars. More recently renewed interest in
hot and dense QED plasmas has been generated due to the possibility of
creating ultra-relativistic electron positron plasmas with high-intensity
lasers ($\approx 10^{18}W/cm^{2}$) [1-3]. Two opposite laser pulses hitting
a thin gold foil can heat up the electrons in the foil up to several MeV $%
(\sim 10^{10}K$ $)$.

The particles propagating in vacuum can be assumed to be the ones with
interactions switched off. When these particles propagate through a medium,
several kinds of interaction processes take place. This makes the properties
of the system different from that in which all\ the particles are assumed to
be completely independent of each other, behaving as freely propagating bare
particles. When dealing with finite temperature environments in QED, where
the particles propagate in statistical background at energies around the
thresholds for particle antiparticle pair production, the temperature
effects need to be appropriately taken into account. These effects arise due
to continuous electron and photon exchanges between particles during the
physical interactions that take place in a heat bath containing hot
particles and antiparticles. The net statistical effects of the background
electrons and photons enter the theory through the fermion and boson
distributions respectively. Finite temperature calculations also provide a
guideline to estimate the density corrections, through chemical potential
effects of the background plasma, for higher order loop corrections.

The thermal background effects are included through the radiative
corrections [4,5]. Self energies and the wavefunctions of the propagating
particles acquire temperature corrections in this environment due to
exchanges of energy and momentum with real particles. The exact state of all
these background particles is unknown since they continually fluctuate
between different configurations. Temperatures of interest in such a
situation are in the range of a few MeV. Thermal propagators in real time
formalism comprise of temperature dependent terms added to the particle
propagators in vacuum theory [6]. In finite temperature electrodynamics,
electric fields are screened due to such interactions.

We use the real time formulation [7] for calculations of wavefunction
renormalization as a second order perturbative correction in $\alpha $ due
to the ease of obtaining the temperature corrections as additive terms to
the usual contribution in vacuum. Here we prefer that the loops with
temperature dependent momenta are integrated before temperature independent
variables in the relevant order $\alpha ^{2}$\ loops in QED, and therefore,
review the electron self energy in Ref. [8,9] earlier. This makes the
calculations of integrations over loop momenta much more simpler and easier
to handle. The results are obtained in a generalized form such that
intermediate temperatures $T\sim m$ are also included while the ranges of
high temperature $T>>m$ and low temperature $T<<m$, are retrieved\ from them
as the limiting cases.

In literature, the ways to compute finite temperature effects on
phase-space, vertex, mass corrections and photon emission or absorption
[10-19 ] are extensively discussed. The finite temperature wave function
renormalization has been dealt with several approaches [12--21],
specifically in the context of weak decay rates during primordial
nucleosynthesis. They agree on using finite temperature Dirac spinors to
obtain the corresponding effective projection operator. Differences in the
spinors presented in Refs. [20] and [21] were also pointed out [22].
However, their results for the case of $\beta $-decay and related processes
agreed with the ones which can be obtained using the approaches that had
already existed for wave function renormalization, except for the case of a
scalar boson decay in fermion-antifermion pair. We work out the two loop
corrections to the finite temperature wave function renormalization in the
generalized temperature framework, for the first time. Section 2 is based on
the re-examined and simplified calculations of loop correction upto two
orders in $\alpha $ that contribute to electron self energy in this
background. The expression for the relative correction in electron mass is
redone and the wave function renormalization constant is calculated in
section 3. Section 4 gives discussion of the results.

\section{Loop corrections to electron self energy\qquad \qquad \qquad \qquad}

At the one loop level, Feynman diagrams are calculated in the usual way by
substituting the finite temperature electron, positron and photon
propagators in place of those in vacuum. In real time formalism, the finite
temperature terms remain separate at order $\alpha $ since the terms
depending on temperature (hot) are additive to temperature independent
(cold) terms in the propagators. Therefore, at the one loop level the hot
and cold loop momenta are integrated separately.

Due to interactions with the background, the electron and positron masses
are known to get enhanced at one-loop and higher loop levels [4-9]. The
photons also acquire dynamically generated mass due to plasma screening
effect [23, 24, 27]. The presence of effective mass implies the fact that
the propagating particles constantly interact with the background.The
radiatively generated thermal mass creates a mass shift and in physical
quantities this acts as a kinematical cut-off, e.g., while determining the
production rate of particles in the heat bath.

Higher order loop corrections are required to study the perturbative
behavior at finite temperature. The two loop integrals comprise a
combination of cold and hot momenta which appear due to an overlap of
temperature dependent and temperature independent terms from the particle
propagators. The loop integrations involve an overlap of finite and
divergent terms due to which these become analytically more complicated. In
such situations, a prefered approach needs to be adopted for integrating
overlapping hot and cold loop momenta, i.e., to integrate over hot loop
momenta before integrating over cold ones, even at the two loop level [24].
This not only helps to simplify the loop integrations but allows one to
handle the statistical effects more appropriately.

The problem of renormalization in finite temperature field theories is
somewhat different from that at zero temperature due to the additional hot
infrared divergences at finite temperatures. The temperature, however, acts
as a regularization parameter for the hot ultraviolet divergences. The
infrared divergences introduced in this framework are also appropriately
removable in particle decay processes via bremstrahlung emission and
absorption effects [5, 25]. This was studied in detail at order $\alpha $
for all the possible ranges of temperature valid in QED including $T\sim m$
[25, 27]. The renormalization of QED\ was also established at the one loop
level for all the relevant ranges in temperatures and chemical potential
[23, 25--27]. \qquad \qquad \qquad \qquad \qquad \qquad \qquad \qquad

\[
\FRAME{itbpF}{1.9692in}{0.9556in}{0in}{}{}{Figure}{\special{language
"Scientific Word";type "GRAPHIC";maintain-aspect-ratio TRUE;display
"USEDEF";valid_file "T";width 1.9692in;height 0.9556in;depth
0in;original-width 2.5832in;original-height 1.2393in;cropleft "0";croptop
"1";cropright "1";cropbottom "0";tempfilename
'LJSD6I00.wmf';tempfile-properties "XPR";}}
\]%
\[
\text{{\small Fig. 1. Two loop electron self energy diagrams and counter
terms }}
\]%
The electron self energy is once again calculated here at the two loop
level, from the point of view of renormalization, by integrating over loop
momenta in the electron self energy diagrams in Fig. 1(a) and 1(b). The
relevant counter terms required to cancel the divergences in these two loop
self energies are included in Fig. 1(c) and 1(d). Removal of overlapping
divergences has been even checked in QED up to the two loop level for $T\sim
m$, $T<<m$ and $T>>m\ $[9]. The integrations over the temperature dependent
momenta are re-examined here and wherever needed are re-done before
temperature independent momentum variables in loops for all the ranges of
temperature that are relevant in QED. The overlapping loops in Fig.1(a)
gives nonzero real contribution to be%
\begin{eqnarray}
\Sigma ^{a}(p) &=&\frac{\alpha ^{2}}{2}{\LARGE [}\frac{1}{2\pi ^{3}}{\Large %
\{}\frac{1}{\varepsilon }[2\NEG{I}-(\NEG{p}+6m)I_{A}+3(2m-\NEG{p})J_{A}+2\NEG%
{J}_{B}]-3(\NEG{p}+4m)I_{A}  \nonumber \\
&&+4\NEG{I}+(12m-5\NEG{p})J_{A}+2\NEG{J}_{B}{\Large \}}+(-1)^{r+1}e^{-r\beta
E}\dsum\limits_{n,r,s=1}^{\infty }{\Large \{[}\frac{3T^{2}}{4}{\LARGE \{}%
f_{+}(n,r)  \nonumber \\
&&\times {\Large [}mf_{+}(s,r){\Large (}\frac{\mathbf{\gamma }.\mathbf{p}}{|%
\mathbf{p}|^{2}}{\Large )}^{2}-f_{-}(s,r)h_{-}(p,\gamma )\frac{\mathbf{%
\gamma }.\mathbf{p}}{|\mathbf{p}|^{2}}{\Large ]}-f_{-}(n,r){\Large [}%
f_{+}(s,r)h_{-}(p,\gamma )\frac{\mathbf{\gamma }.\mathbf{p}}{|\mathbf{p}|^{2}%
}  \nonumber \\
&&-\frac{1}{m}{\Large \{}1-\frac{2}{3}\gamma ^{0}\frac{\mathbf{\gamma }.%
\mathbf{p}}{v|\mathbf{p}|}+{\Large (}\frac{\mathbf{\gamma }.\mathbf{p}}{v|%
\mathbf{p}|}{\Large )}^{2}{\Large \}}f_{-}(s,r){\Large ]}{\LARGE \}}+2T%
{\Large [}(4+\NEG{p}\frac{\mathbf{\gamma }.\mathbf{p}}{|\mathbf{p}|^{2}}%
)f_{+}(s,r)  \nonumber \\
&&+\NEG{p}h_{-}(p,\gamma ){\Large \{}f_{-}(s,r)-f_{-}(n,r)\frac{1}{m}\frac{%
I_{C}}{8\pi }{\Large \}}+\frac{I_{B}}{8\pi }{\Large \{(}4-\NEG{p}\frac{%
\mathbf{\gamma }.\mathbf{p}}{|\mathbf{p}|^{2}}{\Large )}f_{+}(n,r)  \nonumber
\\
&&+\frac{{\Large (}\NEG{p}+m{\Large )}I_{C}}{8\pi }{\Large \}]}{\LARGE \}}%
+(-1)^{s}{\LARGE [}T^{2}{\Large \{}f_{+}(n,r){\Large [}\frac{\mathbf{\gamma }%
.\mathbf{p}}{|\mathbf{p}|^{2}}{\Large (}1-3m\frac{\mathbf{\gamma }.\mathbf{p}%
}{|\mathbf{p}|^{2}}{\Large )}f_{+}(s,r)  \nonumber \\
&&+3h_{-}(p,\gamma )\frac{\mathbf{\gamma }.\mathbf{p}}{|\mathbf{p}|^{2}}%
f_{-}(s,r){\Large ]}-{\Large [\{}\frac{1}{m}h_{-}(p,\gamma )-3m{\Large (}%
\frac{\mathbf{\gamma }.\mathbf{p}}{|\mathbf{p}|^{2}}{\Large )}^{2}{\Large \}}%
f_{+}(s,r)  \nonumber \\
&&-\frac{3}{m}h_{+}(p,\gamma )f_{-}(s,r){\Large ]}f_{-}(n,r){\Large \}}-T%
{\Large \{}{\LARGE [}{\Large \{(}5\NEG{p}+3m^{2}\frac{\mathbf{\gamma }.%
\mathbf{p}}{|\mathbf{p}|^{2}}{\Large )}\frac{\mathbf{\gamma }.\mathbf{p}}{|%
\mathbf{p}|^{2}}-12{\Large \}}f_{+}(n,r)  \nonumber \\
&&+\frac{5\NEG{p}}{m}h_{-}(p,\gamma )f_{-}(n,r){\LARGE ]}\func{Ei}_{-}-%
{\Large [}\frac{1}{m}f_{-}(n,r)h_{+}(p,\gamma )-\frac{\mathbf{\gamma }.%
\mathbf{p}}{|\mathbf{p}|^{2}}f_{+}(n,r)h_{-}(p,\gamma ){\Large ]}  \nonumber
\\
&&\times 3E\func{Ei}_{+}+{\Large \{[}3h_{-}(p,\gamma )\frac{\mathbf{\gamma }.%
\mathbf{p}}{v^{2}}f_{+}(n,r)-\frac{1}{m}f_{-}(n,r)\{3|\mathbf{p}%
|^{2}h_{+}(p,\gamma )  \nonumber \\
&&-5E\NEG{p}h_{-}(p,\gamma )\}{\Large ][}\frac{2e^{-rm\beta }}{m}\sinh
sm\beta +\frac{(r\func{Ei}_{+}+s\func{Ei}_{-})}{T}{\Large ]}-{\Large [}\frac{%
(r\func{Ei}_{+}-s\func{Ei}_{-})}{T}  \nonumber \\
&&+\frac{2e^{-rm\beta }}{m}\cosh sm\beta {\Large ]}m{\Large [\{}%
h_{-}(p,\gamma )-3m{\Large (}\frac{\mathbf{\gamma }.\mathbf{p}}{|\mathbf{p}%
|^{2}}{\Large )}^{2}{\Large \}}f_{-}(n,r)-\frac{m}{2}\frac{\mathbf{\gamma }.%
\mathbf{p}}{|\mathbf{p}|^{2}}  \nonumber \\
&&\times {\Large (}1-3m\frac{\mathbf{\gamma }.\mathbf{p}}{|\mathbf{p}|^{2}}%
{\Large )}f_{+}(n,r){\Large ]}-{\Large [}2\gamma ^{0}T{\LARGE \{}1+\frac{T}{m%
}f_{+}(s,r){\LARGE \}}+\NEG{p}T{\LARGE \{}\frac{2}{m}h_{-}(p,\gamma ) 
\nonumber \\
&&-\frac{\mathbf{\gamma }.\mathbf{p}}{|\mathbf{p}|^{2}}{\Large \}]}%
f_{-}(s,r)+\{E\func{Ei}_{+}-m\gamma ^{0}\func{Ei}_{-}\}+\NEG{p}{\Large \{}%
\frac{m^{2}}{2}\frac{\mathbf{\gamma }.\mathbf{p}}{|\mathbf{p}|^{2}}+\frac{%
2E^{2}}{m}h_{-}(p,\gamma ){\Large \}}  \nonumber \\
&&\times {\Large \{}\frac{e^{-m\beta (s+r)}-e^{-m\beta (r-s)}}{m}+\beta (r%
\func{Ei}_{+}-s\func{Ei}_{-}){\Large \}]}\frac{I_{C}}{8\pi }{\Large ]\}].}
\end{eqnarray}%
where%
\[
f_{\pm }(n,r)={\Large \{}\frac{1}{(n+r)}\pm \frac{1}{(n-r)}{\Large \}};\
f_{\pm }(s,r)={\Large \{}\frac{1}{(s+r)}\pm \frac{1}{(s-r)}{\Large \}},
\]%
\[
h_{\pm }(p,\gamma )=(\gamma ^{0}\pm \frac{\mathbf{\gamma }.\mathbf{p}}{v|%
\mathbf{p}|}),
\]%
\[
\func{Ei}_{\pm }=\func{Ei}[-m\beta (r+s)]\pm \func{Ei}[-m\beta (r-s),
\]%
\[
I_{A}=8\pi \int_{0}^{\infty }\frac{dk}{k}n_{B}(k),
\]%
\[
I_{B}=8\pi \dsum\limits_{r=1}^{\infty }(-1)^{r}\int_{0}^{\infty }\frac{dk}{k}%
e^{-r\beta (p-k)}n_{B}(k),
\]%
\[
I_{C}=8\pi \dsum\limits_{r=1}^{\infty }(-1)^{r}\int_{0}^{\infty }\frac{dk}{k}%
e^{-r\beta k}n_{B}(k),
\]%
\[
J^{A}\simeq -8\pi b(m\beta ),
\]%
\[
\frac{I^{0}}{E}=-\frac{2\pi ^{3}T^{2}}{3E^{2}v}\ln \frac{1-v}{1+v},
\]%
\[
\frac{\mathbf{I.p}}{\mathbf{p}^{2}}=-\frac{2\pi ^{3}T^{2}}{3E^{2}v^{3}}\{\ln 
\frac{1-v}{1+v}+2v\},
\]%
\[
\frac{J_{B}^{0}}{E}\simeq 4\pi \lbrack \frac{T}{\mathbf{p}E}\ln \frac{1+v}{%
1-v}\{ma(m\beta )-Tc(m\beta )\}-3b(m\beta )],
\]%
\[
\frac{\mathbf{J_{B}.p}}{\mathbf{p}^{2}}\simeq \frac{\pi }{v^{2}E^{2}}%
[\{E^{2}-\frac{2}{3}m^{2}\}b(m\beta )+4T\{\frac{1}{v}\ln \frac{1+v}{1-v}%
+2\}\{ma(m\beta )-Tc(m\beta )\}],
\]%
with $v=\frac{|\mathbf{p}|}{p_{_{0}}},$ $(p_{_{0}}=E)$,%
\[
a(m\beta )=\ln (1+e^{-m\beta }),
\]%
\[
b(m\beta )=\dsum\limits_{_{n=1}}^{\infty }(-1)^{n}\func{Ei}(-nm\beta ),
\]%
\[
c(m\beta )=\dsum\limits_{_{n=1}}^{\infty }(-1)^{n}\frac{e^{-nm\beta }}{n^{2}}%
,
\]%
\ and $\func{Ei}(-x)$ is the error integral given by%
\[
\func{Ei}(-x)=-\int_{x}^{\infty }\frac{dt}{t}e^{-t}.
\]%
The non vanishing real terms from loop within loop correction in Fig.1(b)
are also re-examined and recalculated, wherever needed, by retaining the\
specific order of integrating loop momenta, i.e., on the integration\ over
the variables in hot momenta before the cold momenta, we get%
\begin{eqnarray}
\Sigma _{\beta }^{b}(p) &=&\alpha ^{2}{\LARGE \{}\frac{2T^{2}}{3m^{2}}(\NEG%
{p}-m)+mT\dsum\limits_{n,r,s=1}^{\infty }(-1)^{s+r}{\LARGE [}{\Large \{}%
\beta (r+s)\func{Ei}[-m\beta (r+s)]  \nonumber \\
&&\text{ \ }+\frac{e^{-m\beta (s+r)}}{m}{\Large \}\{}\frac{1}{(n-r)}[\frac{2E%
}{m}+\gamma ^{0}(\frac{1}{2}-\frac{E^{2}}{m^{2}})+r\beta (m-\frac{\NEG{p}}{2}%
)]\text{\ \ \ \ \ \ \ \ \ \ \ \ \ }\   \nonumber \\
&&+\frac{1}{(n-s)}{\Large [}h(p,\gamma )+m\frac{\mathbf{\gamma }.\mathbf{p}}{%
|\mathbf{p}|^{2}}{\Large ]\}}+\frac{1}{2(n-r)}{\Large \{}\frac{2}{m}-\frac{%
E\gamma ^{0}}{m^{2}}  \nonumber \\
&&+r\beta \lbrack h(p,\gamma )+m\frac{\mathbf{\gamma }.\mathbf{p}}{|\mathbf{p%
}|^{2}}]\func{Ei}[-m\beta (r+s)]{\Large \}}{\LARGE ]}  \nonumber \\
&&-\frac{\pi T}{6|\mathbf{p}|}\dsum\limits_{n,r,s=1}^{\infty
}(-1)^{r+1}e^{-n\beta E}{\LARGE [}h(p,\gamma ){\Large \{[}1+(-1)^{s}]\frac{%
e^{-m\beta (r-n-s)}}{r-n-s}{\Large \}}  \nonumber \\
&&-T\frac{\mathbf{\gamma }.\mathbf{p}}{|\mathbf{p}|^{2}}\frac{e^{-m\beta
(r-n)}}{r-n}+{\Large (}2-\frac{m\mathbf{\gamma }.\mathbf{p}}{|\mathbf{p}|^{2}%
}{\Large )\{}m\beta (r-n)\func{Ei}[-m\beta (r-n)]  \nonumber \\
&&-e^{-m\beta (r-n)}+\beta \lbrack 1+(-1)^{1+s}]\func{Ei}[-m\beta (r-n-s)]%
{\Large \}}{\LARGE ]\}.}
\end{eqnarray}%
In Eqs. (1) and (2), the preferred order of integration not only
sufficiently eases the calculations but the results are also simpler as
compared to those in Ref. [9] where this preference was not realized. The
electron self energies at the two loop level in Fig 1(a) and 1(b) are then
combined and rearranged to obtain the temperature corrections to the
electron mass and wavefunction.

\section{The Wavefunction Renormalization}

To incorporate finite temperature effecst on physical processes beyond the
tree level, one needs to have a consistent method of temperature dependent
renormalization. As already mentioned, renormalizability of the electron
mass was done through the order by order cancellation of singularities up to
two loop level. It can be easily checked that the second order\ in $\alpha $
correction is much smaller than the first order contribution so that the
perturbative behavior is valid. In a background, with $T\neq 0$, the Lorentz
invariance is broken and momentum independent renormalization constant is no
longer sufficient. Donoghue and Holstein used the temperature dependent
propagator to modify electron mass as well as the spinors accordingly [5].

The shift in the electron mass due to finite temperature effects is
calculated here from Eqs. (1) and (2). For this all the finite terms in
electron self energy upto second order in $\alpha $ are put together.
Following Ref. [5] the physical mass of the electron at one loop was
obtained in Ref. [25]\ in generalized form, by writing%
\[
\Sigma (p)=A(p)E\gamma _{_{0}}-B(p)\vec{p}.\vec{\gamma}-C(p), 
\]%
where $A(p)$, $B(p)$, and $C(p)$ are the relevant coefficients. Taking the
inverse of the propagator with momentum and mass term separated as%
\[
S^{-1}(p)=(1-A)E\gamma ^{o}-(1-B)p.\gamma -(m-C), 
\]%
and the physical mass $m_{phy}=m+\delta m^{(1)}+\delta m^{(2)},$ was deduced
by locating the pole of the propagator $\frac{i(\NEG{p}+m)}{%
p^{2}-m^{2}+i\varepsilon }$ . $\delta m^{(1)}$ and $\delta m^{(2)}$\ is the
shift in electron mass due to temperature effects at one and two loop level
respectively.$~$Using the same procedure, the relative shift in electron
mass at the two loop level was obtained [9]. This is recalculated here,
wherever required, and after recombining the similar summations it becomes:%
\begin{eqnarray}
\frac{\delta m^{(2)}}{m} &=&2\alpha ^{2}\dsum\limits_{r=1}^{\infty }{\LARGE [%
}\frac{T^{2}}{m^{2}}{\LARGE \{}\dsum\limits_{n=3}^{r+1}\text{ }(-1)^{n+r+1}%
\frac{\pi m\text{ }}{6|\mathbf{p}|}\frac{e^{-\beta (rE+mn)}}{n}\   \nonumber
\\
&&-\frac{3}{8}(-1)^{r}\frac{e^{-r\beta E}}{|\mathbf{p}|^{2}}{\Large [}\frac{%
9E^{2}}{2m^{2}}+6\dsum\limits_{s=3}^{r+1}\frac{1}{s}+4\dsum%
\limits_{n,s=3}^{r+1}\frac{1}{ns}+(-1)^{s-r}{\Large \{}\frac{9E}{m}%
(3+4\dsum\limits_{s=3}^{r+1}\frac{1}{s}{\Large )}  \nonumber \\
&&+2{\Large (}\frac{|\mathbf{p}|^{2}}{m^{2}}-3{\Large )(}9+18\dsum%
\limits_{s=3}^{r+1}\frac{1}{s}+8\dsum\limits_{n,s=3}^{r+1}\frac{1}{ns}%
{\Large )\}]}+\frac{4}{E^{2}v^{2}}{\LARGE \}}-\frac{m^{2}}{\pi ^{2}}c(m\beta
)  \nonumber \\
&&-\frac{T}{m}{\LARGE \{}\frac{\pi \text{ }}{6|\mathbf{p}|}%
\dsum\limits_{s=2}^{r+1}\dsum\limits_{n=1}^{s+1}\frac{e^{-\beta (rE+mn)}}{n}%
{\LARGE [}1{\LARGE \ }-{\Large \{}(-1)^{r+n}-(-1)^{s+n}{\Large \}}{\LARGE ]}
\nonumber \\
&&+{\Large [\{}\func{Ei}(-m\beta )-\func{Ei}(-2m\beta ){\Large \}\{}\frac{9E%
}{4}{\Large (}\frac{E}{|\mathbf{p}|^{2}}-\frac{1}{m}{\Large )}+{\Large (}%
\frac{5E}{m}-21+\frac{E^{2}}{2m^{2}}{\Large )}\dsum\limits_{n=3}^{r+1}\frac{1%
}{n}{\Large \}}  \nonumber \\
&&+{\Large \{}\frac{9}{4v^{2}}-\dsum\limits_{n=1}^{s+1}\dsum%
\limits_{s=3}^{r+1}{\Large [}1-E^{2}{\Large (}\frac{1}{2m^{2}}+\frac{3}{|%
\mathbf{p}|^{2}}{\Large )}+\frac{3E}{m}{\Large ]\}}(-1)^{s}\func{Ei}%
(-sm\beta )]  \nonumber \\
&&+e^{-rm\beta }{\LARGE \{[}\frac{9E}{2v^{2}}+2{\LARGE (}\frac{3E}{v^{2}}+%
\frac{3|\mathbf{p}|^{2}}{m}-5E{\LARGE )}\dsum\limits_{n=3}^{r+1}\frac{1}{n}%
{\LARGE ]}\dsum\limits_{s=1}^{\infty }\sinh sm\beta  \nonumber \\
&&-\frac{3m^{3}}{|\mathbf{p}|^{2}}{\Large (}\frac{3}{4}-\dsum%
\limits_{n=3}^{r+1}\frac{1}{n}{\LARGE )}\dsum\limits_{s=1}^{\infty }\cosh
sm\beta {\LARGE \}]\}}+{\Large {\LARGE \{}}\frac{9m}{4|\mathbf{p}|^{2}}%
{\Large (}E^{3}+\frac{m^{3}}{2}{\Large )}  \nonumber \\
&&+{\Large [}\frac{3m}{|\mathbf{p}|^{2}}(E^{3}+m^{3})+5mE-3|\mathbf{p}|^{2}%
{\Large ]}\dsum\limits_{n=3}^{r+1}\frac{1}{n}{\LARGE \}}{\Large \{}\func{Ei}%
(-m\beta )-2\func{Ei}(-2m\beta ){\Large \}}  \nonumber \\
&&-\dsum\limits_{n=3}^{r+1}{\LARGE \{}\dsum\limits_{s=1}^{r+1}\frac{(-1)^{s}%
}{n}{\Large [}\frac{m^{2}r}{2}e^{-sm\beta }+{\LARGE \{}s{\LARGE (}2mE-\frac{%
E^{3}}{m}{\LARGE )}+\frac{m^{2}(s-r)}{2}{\LARGE \}}\func{Ei}(-sm\beta )%
{\Large ]}  \nonumber \\
&&-\frac{\pi m^{2}\text{ }}{3|\mathbf{p}|}{\LARGE [}e^{-\beta rE}\text{ }%
(-1)^{n+r}(n+1){\LARGE \ }-\dsum\limits_{s=2}^{r+1}(-1)^{n+s}{\LARGE ]}\func{%
Ei}(-nm\beta ){\LARGE \}].}
\end{eqnarray}

From the reviewed expression for the electron self energy obtained in Eqs.
(1) and (2) the relation for the wave function renormalization constant is
derived. This comes out to be%
\begin{eqnarray}
Z_{2}^{-1} &=&\frac{\partial \Sigma }{\partial \NEG{p}}  \nonumber \\
&=&1-\alpha {\LARGE [}\frac{1}{4\pi }\left( \frac{3}{\varepsilon }-4\right) +%
\frac{5}{\pi }b(m\beta )+\frac{I_{A}}{4\pi ^{2}}  \nonumber \\
&&-\frac{T^{2}}{\pi vE^{2}}\ln \frac{1+v}{1-v}{\LARGE \{}\frac{\pi ^{2}}{6}%
-c(m\beta )+m\beta a(m\beta ){\LARGE \}]}  \nonumber \\
&&-\alpha ^{2}{\LARGE [}\frac{1}{4\pi ^{3}}\{\frac{1}{\varepsilon }%
(I_{A}-3J_{A})+(3I_{A}+5J_{A})\}-\frac{2T^{2}}{3\pi ^{3}m^{2}}  \nonumber \\
&&+\frac{m}{8\pi }\dsum\limits_{n,r,s=1}^{\infty }(-1)^{s+r}{\Large \{}r[%
\frac{e^{-m\beta (s+r)}}{m}-\beta (r+s)\func{Ei}\{-m\beta (r+s)\}]{\Large \}}
\nonumber \\
&&+\frac{1}{8}\dsum\limits_{n,r,s=1}^{\infty }(-1)^{r}T{\LARGE \{}e^{-r\beta
E}{\LARGE [}f_{+}(s,r)\frac{\mathbf{\gamma }.\mathbf{p}}{|\mathbf{p}|^{2}}-%
\frac{I_{B}I_{C}}{64\pi ^{2}}  \nonumber \\
&&+h(p,\gamma )\{f_{-}(n,r)\frac{I_{C}}{8\pi }-f_{-}(s,r)\}+f_{+}(n,r)\frac{%
\mathbf{\gamma }.\mathbf{p}}{|\mathbf{p}|^{2}}\frac{I_{B}}{8\pi }{\LARGE ]} 
\nonumber \\
&&+{\LARGE [\{}5\frac{\mathbf{\gamma }.\mathbf{p}}{|\mathbf{p}|^{2}}%
f_{+}(n,r)-\frac{5}{m}h(p,\gamma )f_{-}(n,r){\LARGE \}}\func{Ei}_{-} 
\nonumber \\
&&+\frac{5E}{m^{2}}h(p,\gamma )f_{-}(n,r){\LARGE \{}\frac{2e^{-rm\beta }}{m}%
\sinh sm\beta +\beta (r\func{Ei}_{+}+s\func{Ei}_{-}){\LARGE \}}  \nonumber \\
&&+{\LARGE \{}\frac{2}{m}h(p,\gamma )-\frac{\mathbf{\gamma }.\mathbf{p}}{|%
\mathbf{p}|^{2}}{\LARGE \}}f_{-}(s,r){\LARGE ]\}}+{\LARGE \{}\frac{m^{2}}{2}%
\frac{\mathbf{\gamma }.\mathbf{p}}{|\mathbf{p}|^{2}}+\frac{2E^{2}}{m}%
h(p,\gamma ){\LARGE \}}  \nonumber \\
&&\times {\LARGE \{}\frac{2e^{-rm\beta }}{m}\sinh sm\beta +\beta (r\func{Ei}%
_{+}-s\func{Ei}_{-}){\LARGE \}}\frac{I_{C}}{8\pi }{\LARGE ].}
\end{eqnarray}%
From this expression for $Z_{2}^{-1},$ not only the behavior at intermediate
temperatures $T\sim m$\ can be extracted but the ranges of high temperature $%
T>>m$, low temperature $T<<m$, can be also retrieved\ from it as limiting
cases.

\section{Results and Discussion}

With the preferred order of integration of hot loops before the cold ones,
the previously calculated self-mass correction terms are redone, wherever
required, for all the possible ranges of temperature. This preference
simplifies the calculations since the statistical effects are taken care of
through hot loop momenta integrations here, before the zero temperature
integration variables are dealt with. Therefore, we have re-written the
electron self energy expressions in QED at the two loop level that were
presented in Ref. [9]. \ This led to the modfied expression for the relative
change in the electron mass at the two loop level in Eq. (3). From these
corrections one can then retrieve the results for all temperature ranges of
interest here, classified as, the high temperature $T>>m$ $($having $m\beta
\longrightarrow 0$ with $e^{-m\beta }$ falling off exponentially as compared
to $\frac{T^{2}}{m^{2}}$), the low temperature $T<<m$ (with fermions
contribution negligible) and the intermediate temperatures $T\sim m$ (by
taking $m\beta \rightarrow 1$).

We calculated here, for the first time in this generalized form, the
wavefunction renormalization constant using the thermal contributions to the
second order self energy diagrams. The divergences get cancelled as usual by
including the counter terms in Fig. 1(c) and 1(d), as already checked [9].
The fermions do not pick any contribution from the heat bath at low
temperature. Therefore, the second order in $\alpha $ corrections to the
electron self energy at low temperature can be retrieved as a limiting case
that contains contribution from hot photons only giving:%
\[
\Sigma _{\beta }(p)\text{ }_{\longrightarrow }^{^{T<<m}}\frac{\alpha ^{2}}{%
4\pi ^{3}}\left[ 4\NEG{I}+\frac{8\pi T^{2}}{3m^{2}}(\NEG{p}-m)\right] . 
\]%
The wave function renormalization constant upto two loops at $T<<m$\ from
Eq. (4) is therefore%
\begin{eqnarray}
&&Z_{2}^{-1}\bigskip \text{ $_{\longrightarrow }^{^{T<<m}}$ }1+\frac{\alpha 
}{4\pi }\left( 4-\frac{3}{\varepsilon }\right) -\frac{\alpha }{4\pi ^{2}}%
\left( I_{A}-\frac{I^{0}}{E}\right)  \nonumber \\
&&\text{ \ \ \ \ \ \ \ \ \ \ \ }-\frac{\alpha ^{2}}{4\pi ^{2}}\left( 3+\frac{%
1}{\varepsilon }\right) I_{A}+\frac{2\alpha ^{2}T^{2}}{3\pi ^{2}m^{2}},
\end{eqnarray}%
which is the same as\ that in Ref. [28]. The high temperature limit for this
constant gives:%
\begin{eqnarray}
&&Z_{2}^{-1}\text{ }_{\longrightarrow }^{^{T>>m}}\text{ }1-\alpha {\Large [}%
\frac{2I_{A}}{\pi }+\frac{1}{4\pi }\left( \frac{3}{\varepsilon }-4\right) +%
\frac{4\pi T^{2}}{3}{\Large ]}  \nonumber \\
&&\text{ \ \ \ \ \ \ \ \ \ \ \ }-\alpha ^{2}{\LARGE [}\frac{1}{4\pi ^{3}}%
{\LARGE \{}\frac{1}{\varepsilon }(I_{A}-3J_{A})+(3I_{A}+5J_{A})-\frac{8T^{2}%
}{3m^{2}}{\LARGE \}}  \nonumber \\
&&\text{ \ \ \ \ \ \ \ \ \ \ \ }+\frac{1}{8}\dsum\limits_{n,r,s=1}^{\infty
}(-1)^{r}T{\LARGE \{}e^{-r\beta E}{\LARGE [}f_{+}(s,r)\frac{\mathbf{\gamma }.%
\mathbf{p}}{|\mathbf{p}|^{2}}-\frac{I_{B}I_{C}}{64\pi ^{2}}  \nonumber \\
&&\text{ \ \ \ \ \ \ \ \ \ \ \ }+h(p,\gamma )\{f_{-}(n,r)\frac{I_{C}}{8\pi }%
-f_{-}(s,r)\}+f_{+}(n,r)\frac{\mathbf{\gamma }.\mathbf{p}}{|\mathbf{p}|^{2}}%
\frac{I_{B}}{8\pi }  \nonumber \\
&&\text{ \ \ \ \ \ \ \ \ \ \ \ }+{\LARGE \{}\frac{2}{m}h(p,\gamma )-\frac{%
\mathbf{\gamma }.\mathbf{p}}{|\mathbf{p}|^{2}}{\LARGE \}}f_{-}(s,r){\LARGE %
]\}].}
\end{eqnarray}

It can be seen from Eq.(7) that the leading contribution in this range of
temperature, $T>>m$, at the two loop level is $\frac{2T^{2}}{3m^{2}}$. The
two loop fermion self energy in QED has been calculated in detail recently
[29] using the hard thermal loop resummation introduced by Braaten and
Pisarski [30]. As far as the renormalization is concerned, the resummed hard
thermal loops (HTL) do not affect it [29]. Hence $Z_{2}$ does not get any
contribution from HTL here. Moreover, it is worth noticing that the thermal
corrections, at second order in $\alpha ,$ to the wavefunction
renormalization constant at extreme temperatures ( $T<<m$ and $T>>m$ ) are
still proportional to $\frac{T^{2}}{m^{2}}$ as in case of the selfmass of
electron. However, the expression for the intermediate temperatures is
significantly different from the self-energy expression obtained earlier
[9]. The calculations around $T\sim m$ for the mass and wavefunction
renormalizations are still cumbersome at the two-loop level, but much less
difficult than those in Ref. [9]. The renormalizability of the theory at
finite temperature can be explicitly checked and holds through the order by
order cancellation of singularities. This provides a platform to include the
general effects due to chemical potential in the hot and dense background,
later on. With the experience of including chemical potential at one loop
level, in real time formulation [23,26,27], it is foreseen that two loop
self energies will be much more complicated but is still worth-doing to
develop a calculational technique for high density hot plasmas or even
superfluids inside the cores of neutron stars[31]. The modified wavefunction
is expected to modify the finite temperature contributions to electroweak
processes [32] as well as the neutrino magnetic moments [33] up to the two
loop level.

\begin{quotation}
\textbf{Acknowledgement}
\end{quotation}

{\small One of the authors (MQH) thanks Higher Education Commission,
Pakistan for providing partial funding under a research grant during this
work.}

\begin{quotation}
\textbf{Figure caption}
\end{quotation}

{\small Fig. 1 \ Two loop electron self energy diagrams and counter terms}

\begin{quotation}
\textbf{References}
\end{quotation}

\begin{enumerate}
\item {\small E.P. Liang, S.C. Wilks, and M. Tabak, Pair production by
ultraintense lasers, Phys. Rev. Lett. 81 (1998) 4887--4890.}

\item {\small B. Shen, J. Meyer-ter-Vehn, Pair and }$\gamma ${\small -photon
production from a thin foil confined by two laser pulses, Phys. Rev. E 65
(2001) 016405--016410.}

\item {\small M.H. Thoma, Ultrarelativistic electron-positron plasma, Eur.
Phys. J. D 55 (2009) 271--278; M.H. Thoma, Colloquium: Field theoretic
description of ultrarelativistic electron positron plasmas, Rev. Mod. Phys.
81 (2009) 959--968.}

\item {\small H.A. Weldon, Covariant calculations at finite temperature: The
relativistic plasma, Phys. Rev. D 26 (1982) 1394--1407.}

\item {\small J.F. Donoghue and B.R. Holstein, Renormalization and radiative
corrections at finite-temperature, Phys. Rev. D 28 (1983) 340--348; ibid 29
(1983) 3004(E).}

\item {\small P. Landsman and Ch G. Weert, Real and imaginary time field
theory at finite temperature and density, Phys. Rep. 145 (1987) 141--249.}

\item {\small V.V. Klimov, Spectrum of elementary Fermi excitations in quark
gluon plasma, Sov. J. Nucl. Phys. 33 (1981) 934--935; L. Dolan and R.
Jackiw, Symmetry behavior at finite temperature, Phys. Rev. D 9 (1974)
3320--3341; S. Weinberg, Gauge and global symmetries at high temperature,
Phys. Rev. D 9 (1974) 3357--3378; C. Bernard, Feynman rules for gauge
theories at finite temperature, Phys. Rev. D 9 (1974) 3312-- 3320. }$\ \ \ $

\item {\small Mahnaz Qader (Haseeb), Samina S. Masood, and K. Ahmed,
Second-order electron mass dispersion relation at finite temperature, Phys.
Rev. D 44 (1991) 3322--3327.}

\item {\small Mahnaz Qader (Haseeb), Samina S. Masood, and K. Ahmed,
Second-order electron mass dispersion relation at finite temperature. II.,
Phys. Rev. D 46 (1992) 5633--5647.}

\item {\small D.A. Dicus, E.W. Kolb, A.M. Gleeson, E.C.G. Sudarshan,
Primordial nucleosynthesis including radiative, Coulomb, and
finite-temperature corrections to weak rates, V.L. Teplitz, and M.S. Turner,
Phys. Rev. D 26 (1982) 2694--2709.}

\item {\small J.-L. Cambier, J.R. Primack, and M. Sher, Finite temperature
radiative corrections to neutron decay and related processes, Nucl. Phys.
B209 (1982) 372--388.}

\item {\small J.F. Donoghue, B.R. Holstein, and R.W. Robinett, Quantum
electrodynamics at finite temperature, Ann. Phys. (N.Y.) 164 (1985) 233--276.%
}

\item {\small A.E. Johansson, G. Peresutti, and B.S. Skagerstam, Quantum
field theory at finite temperature: Renormalization and radiative
corrections, Nucl. Phys. B278 (1986) 324--342.}

\item {\small W. Keil, Radiative corrections renormalization and at
finite-temperature: A real time approach, Phys. Rev. D 40 (1989) 1176--1198.}

\item {\small R. Baier, E. Pilon, B. Pire, and D. Schiff, Finite-temperature
radiative corrections to early universe neutron-proton ratio: Cancellation
of infrared and mass singularities, Nucl. Phys. B336 (1990) 157--183.}

\item {\small W. Keil and R. Kobes, Mass and wave function renormalization
at finite temperature, Physica A 158 (1989) 47--57.}

\item {\small M. LeBellac and D. Poizat, Renormalization of external lines
in relativistic field theories at finite temperature, Z. Phys. C 47 (1990)
125--132.}

\item {\small T. Altherr and P. Aurenche, Fermion self-energy corrections in
perturbative theory at finite temperature, Phys. Rev. D 40 (1989) 4171--4177.%
}

\item {\small R.L. Kobes and G.W. Semeneff, Discontinuity of Green functions
in field theory at finite temperature and density (I), Nucl. Phys. B260
(1985) 714--746; Discontinuity of Green functions in field theory at finite
temperature and density (II), B272 (1986) 329--364.}

\item {\small R.F. Sawyer, Temperature-dependent wave function
renormalization and weak interaction rates prior to nucleosynthesis, Phys.
Rev. D 53 (1996) 4232--4236.}

\item {\small I.A. Chapman, Finite temperature wave-function
renormalization, a comparative analysis, Phys. Rev. D 55 (1997) 6287--6291.}

\item {\small S. Esposito, G. Mangano, G. Miele, and O. Pisanti, Wave
function renormalization at finite temperature, Phys. Rev. D 58 (1998)
105023-1--5.}

\item {\small K. Ahmed and Samina S. Masood, Vacuum polarization at finite
temperature and density in QED, Ann. Phys. (N.Y.) 207 (1991) 460--473.}

\item {\small Samina S. Masood and Mahnaz Q. Haseeb, Second-order
corrections to QED coupling at low temperature, Int. J. of Mod. Phys. A 23
(2008) 4709--4719.}

\item {\small K. Ahmed and Samina Saleem (Masood), Renormalization and
radiative corrections at finite-temperature reexamined, Phys. Rev. D 35
(1987) 1861--1871.}

\item {\small K. Ahmed and Samina Saleem (Masood), Finite-temperature and
-density renormalization effects in QED, Phys. Rev. D 35 (1987) 4020--4023.}

\item {\small Samina S. Masood, Photon mass in the classical limit of
finite-temperature and -density QED, Phys. Rev. D 44 (1991) 3943--3948;
Samina S. Masood, Renormalization of QED in superdense media, Phys. Rev. D\
47 (1993) 648--652.}

\item {\small Mahnaz Q. Haseeb and Samina S. Masood, Two loop low
temperature corrections to electron self energy, Chin. Phys. C, published
online Nov. 29, 2010 (in press).}

\item {\small E. Mottola and Z. Sz\'{e}p, Systematics of high temperature
perturbation theory: The two-loop electron self energy in QED, Phys. Rev. D
81 (2010) 025014--025052.}

\item {\small E. Braaten and R.\thinspace D. Pisarski, Soft amplitudes in
hot gauge theories: A general analysis, Nucl. Phys. B 337 (1990) 569--634;
E. Braaten and R.\thinspace D. Pisarski, Calculation of the gluon damping
rate in hot QCD, Phys. Rev. D 42 (1990) 2156--2160; R.\thinspace D.
Pisarski, Scattering amplitudes in hot gauge theories, Phys. Rev. Lett. 63
(1989) 1129--1132.}

\item {\small D. Page, et. al., Rapid cooling of the neutron star in
cassiopeia a triggered by neutron superfluidity in dense matter, Phys. Rev.
Lett. 106 (2011) 081101-081108. }

\item {\small Mahnaz Qader (Haseeb) and Samina S. Masood, Finite-temperature
and -density corrections to electroweak processes, Phys. Rev. D 46 (1992)
5110--5116. }

\item {\small Samina S. Masood, Neutrino physics in hot and dense media,
Phys. Rev. D 48 (1993) 3250--3258.}
\end{enumerate}

\end{document}